\documentclass[useAMS,usenatbib]{mn2e}
\usepackage{graphicx,epsfig}

\def\asec{\ifmmode ^{\prime\prime}\else$^{\prime\prime}$\fi}

\def\msunyr{\mbox{\,${\rm M_{\odot}\, yr^{-1}}$}}
\def\mdot{\dot M}

\def\degs{\ifmmode ^{\circ}\else$^{\circ}$\fi}
\def\amin{\ifmmode ^{\prime}\else$^{\prime}$\fi}
\def\asec{\ifmmode ^{\prime\prime}\else$^{\prime\prime}$\fi}
\def\fdg{\hbox{$.\!\!^\circ$}}          
\def\farcs{\hbox{$.\!\!^{\prime\prime}$}}  

\def\degs{\ifmmode ^{\circ}\else$^{\circ}$\fi}
\def\amin{\ifmmode ^{\prime}\else$^{\prime}$\fi}

\def\EE#1{\times 10^{#1}}
\def\cm{\mbox{\,cm}}

\def\cm3{\mbox{\,cm$^{-3}$}}
\def\kms{\mbox{\,km~s$^{-1}$}}

\def\erg{\mbox{\,ergs}}
\def\ergs{\mbox{\,ergs~s$^{-1}$}}

\def\kms{\mbox{\,km s$^{-1}$}}

\def\lsim{\!\!\!\phantom{\le}\smash{\buildrel{}\over
 {\lower2.5dd\hbox{$\buildrel{\lower2dd\hbox{$\displaystyle<$}}\over
                                 \sim$}}}\,\,}
\def\gsim{\!\!\!\phantom{\ge}\smash{\buildrel{}\over
{\lower2.5dd\hbox{$\buildrel{\lower2dd\hbox{$\displaystyle>$}}\over
                               \sim$}}}\,\,}

\title{High-resolution observations of SN~2001gd in NGC~5033}
\author[M.A.\ P\'erez-Torres et al.]
{M.A.\ P\'erez-Torres$^1$\thanks{E-mail: torres@iaa.es}, 
   A.\ Alberdi$^1$,
   J.M.\ Marcaide$^2$, 
   M.A.\ Guerrero$^1$,  
   P.\ Lundqvist$^3$,
   \newauthor   
   I.I.\ Shapiro$^4$,
   E.\ Ros$^5$,
   L.\ Lara$^{6,1}$,
   J.C.\ Guirado$^2$,
   K.W.\ Weiler$^7$,
   C.J.\ Stockdale$^7$ \\
$^1$Instituto de Astrof\'{\i}sica de Andaluc\'{\i}a, CSIC, Apdo.
Correos 3004, E-18080 Granada, Spain \\
$^2$Departamento de Astronom\'{\i}a y Astrof\'{\i}sica, 
Universidad de Valencia, E-46100 Burjassot, Spain \\
$^3$Stockholm Observatory, AlbaNova, Roslagstullsbacken 21,
SE-10691 Stockholm, Sweden\\
$^4$Harvard-Smithsonian Center for Astrophysics, 
60 Garden St. MS 51, Cambridge, MA 02138, USA\\
$^5$Max-Planck-Institut f\"ur Radioastronomie, Auf dem H\"ugel 69,
D-53121 Bonn, Germany \\
$^6$Departamento de F\'{\i}sica Te\'orica y del Cosmos, 
Universidad de Granada, E-18071 Granada, Spain\\
$^7$Naval Research Laboratory, Code 7213,
Washington, DC 20375-5320, USA  \\
}

\date{Accepted 2005 April 11.
      Received 2005 April  7;
      in original form 2004 November 10
     }

\pagerange{\pageref{firstpage}--\pageref{lastpage}} \pubyear{2005}
\def\LaTeX{L\kern-.36em\raise.3ex\hbox{a}\kern-.15em
T\kern-.1667em\lower.7ex\hbox{E}\kern-.125emX}

\begin{document}
\label{firstpage}
\maketitle

\begin{abstract}
We report on 8.4 GHz VLBI observations of SN~2001gd in the spiral galaxy
NGC~5033 made on 26 June 2002 (2002.48) and 8 April 2003 (2003.27).
We used the interferometric visibility data to estimate angular diameter sizes for 
the supernova by model fitting. 
Our data nominally suggests a relatively strong deceleration for the 
expansion of SN~2001gd, but we cannot dismiss the possibility of a free supernova expansion. 
From our VLBI observations on 8 April 2003, 
we inferred a minimum total energy in relativistic particles and magnetic fields 
in the supernova shell of
$E_{\rm min}$=(0.3--14)$\,\times\, 10^{47}\,$ ergs, and
a corresponding equipartition average magnetic 
field of $B_{\rm min}$=\,(50--350)\,mG. 
We also present multiwavelength VLA measurements of SN~2001gd
made at our second VLBI epoch at the
frequencies of 1.4, 4.9, 8.4, 15.0, 22.5, and 43.3 GHz.
The VLA data are well fit by an optically thin, synchrotron
spectrum ($\alpha = -1.0 \pm 0.1; S_\nu\, \propto\, \nu^{\alpha})$,
partially absorbed by thermal plasma.
We obtain a supernova flux density of (1.02$\pm$0.05)\,mJy at the
observing frequency of 8.4 GHz for the second epoch, 
which results in an isotropic radio luminosity of
(6.0$\pm$0.3)$\times\,10^{36}$\ergs\, between 1.4 and 43.3\,GHz,
at an adopted distance of 13.1~Mpc.
Finally, we report on an {\it XMM-Newton} X-ray detection of SN~2001gd
on 18 December 2002.
The supernova X-ray spectrum is consistent with optically thin emission
from a soft component (associated with emission from 
the reverse shock) at a temperature around 1 keV.
The observed flux corresponds to an isotropic X-ray luminosity of 
$L_{\rm X}=(1.4\pm0.4)\EE{39}$ ergs~s$^{-1}$ in the 0.3--5 keV band.
We suggest that both radio and X-ray observations of SN~2001gd indicate that 
a circumstellar interaction similar to that displayed by SN~1993J in M~81
is taking place.
\end{abstract}


\section{Introduction}\label{sec,intro}

SN~2001gd in NGC~5033 was discovered by \citet{nakano01} 
on 24.82 November 2001;
its explosion date is uncertain.
The supernova had a visual magnitude then of 14.5, and was located
$\sim$3' north-northwest of the nucleus of NGC~5033.
\citet{nakano01} reported the following position
for SN~2001gd: $\alpha$=13h13m23s.89, $\delta$=+36$^\circ$38'17".7
(equinox J2000.0).
They also reported that there was no star visible at the
above position on earlier frames taken between 1996 and April 2001.
An optical spectrum obtained by P. Berlind on 4.52 December 2001 \citep{matheson01},
showed SN~2001gd to be a Type IIb supernova well past maximum light.
\citet{matheson01} pointed out that the spectrum was almost identical
to that of SN~1993J obtained on day 93 after explosion \citep{matheson00}.
Since SN~1993J was a strong radio emitter, it was natural to expect SN~2001gd
would also be a strong radio emitter.
\citet{stockdale02} detected SN~2001gd on 8 February 2002
at cm-wavelengths with the Very Large Array (VLA).
Their continual monitoring of SN~2001gd since its first radio detection
has confirmed that SN~2001gd is similar in
its radio properties to SN~1993J (peak luminosity at 6~cm, 
time to 6~cm peak, spectral index; \citealt{stockdale03}),
as in its optical properties.
Given the similarities between the optical spectrum of SN~2001gd on 4.52 December
2001 and that of SN~1993J at an age of 93 days \citep{matheson01}, 
we assume throughout our paper that the supernova exploded 
93 days before 5 December, therefore fixing that event at 
$t_0$\,=\,3 September 2001.

A preliminary report on our radio findings is presented in \citet{mapt05}. 
Here, we present the results of our two-epoch VLBI observing campaign on SN~2001gd,
complemented with VLA observations and {\it XMM-Newton} archival X-ray data.
The paper is organized as follows: we report on our VLBI and VLA
radio observations, and on {\it XMM-Newton} archival X-ray data
in Sect.~\ref{sec,obs}; 
we present our VLBI images of SN~2001gd in Sect.~\ref{sec,vlbi-images};
we present and discuss our results in Sect.~\ref{sec,discussion}. 
We summarize our main conclusions in Sect.~\ref{summary}.
We assume throughout our paper a distance of 13.1~Mpc to 
the host galaxy of SN~2001gd, NGC~5033, based on its redshift
($z=0.002839$; \citealt{falco99}) and assumed
values of $H_0 = 65\kms$~Mpc$^{-1}$ and $q_0$=0. 
At this distance, 1~mas corresponds to a linear size of 0.063~pc.

\section{Observations and data reduction}
\label{sec,obs}

\subsection{VLBI measurements}
\label{subsec,vlbi-obs}

We observed SN~2001gd on 26 June 2002 (2002.48) at a frequency of 8.4 GHz,
using a VLBI array that included the following 12 antennas
(diameter, location): 
The VLBA (25~m, 10 identical antennas across the USA),
Green Bank (100~m, WV, USA), and Effelsberg (100~m, Germany).
We also observed SN~2001gd on 8 April 2003 (2003.27) at the same frequency
with the above array, complemented with the 
phased-VLA (130~m-equivalent, NM, USA), and Medicina and Noto (32~m each, Italy).
However, for our second epoch we had to discard three 
antennas (Hancock, North Liberty, and Noto) due to the poor 
quality of their data. 
The duration of each observing run was eight hours.
The telescope systems recorded right-hand circular polarization (RCP),
and used a bandwidth of 64\,MHz, except for the VLA (bandwidth of 50\,MHz).
The data were correlated at the VLBA Correlator of the National
Radio Astronomy Observatory (NRAO) in Socorro (NM, USA).
The correlator used an averaging time of 2 s.

Since previous VLA observations of SN\,2001gd showed it to be fainter
than 6~mJy at 8.4~GHz, our VLBI observations were carried out in 
phase-reference mode.
SN\,2001gd and the nearby International Celestial Reference Frame 
(ICRF) source J1317+34 ($z=1.050$; \citealt{hb89})
were alternately observed through each eight-hour long VLBI run.
The observations consisted of $\sim145$\, s scans
on SN~2001gd
and $\sim95\,$ s scans on J1317+34,
plus $\sim60$ additional seconds of antenna slew time to make
a duty cycle of $\sim$300 s.
In each observing run, the total on-source time
was $\approx$4.2\,h and $\approx$2.5\,h
for SN~2001gd and J1317+34, respectively.
J1317+34 was also used as the amplitude calibrator for SN\,2001gd.
The sources 4C~39.25 and J1310+32 (both ICRF sources) were observed as fringe finders.

The correlated data were analyzed using the Astronomical Image
Processing System ({\it AIPS}).
The visibility amplitudes were calibrated using the system temperature
and gain information provided for each telescope.
The instrumental phase and delay offsets among the 8~MHz baseband
converters in each antenna were corrected using a phase calibration
determined from observations of 4C\,39.25 and J1310+32, for our
VLBI observations on 26 June 2002 and 8 April 2003, respectively.
We fringe-fit the data for the calibrator J1317+34
in a standard manner; these data were then 
exported to the Caltech program DIFMAP
(Shepherd et al. 1995) for imaging purposes.
The final source image obtained for J1317+34 was then included as an 
input image in a new round of fringe-fitting for J1317+34 within {\it AIPS}.
In this way, the results obtained for the phase delays and delay-rates for
J1317+34 were nearly structure-free.
These new values were then interpolated and applied 
to the source SN\,2001gd using {\it AIPS} standard procedures.
The SN\,2001gd data were then transferred into the DIFMAP program.
We used standard phase self-calibration techniques using a time interval of 
3 minutes, to obtain the images shown in Sect.~\ref{sec,vlbi-images}.

\subsection{VLA measurements}
\label{subsec,vla-obs}

Our second observing epoch (2003.27) included the phased-VLA (in D-configuration)
as an element of our VLBI array.
We allocated about one hour of the VLA time to determine the radio spectrum of SN~2001gd.
We observed at 1.4, 4.9, 8.4, 15.0, 22.5, and 43.3 GHz,
in standard continuum mode.
We observed in both senses of circular polarization and
each frequency band was split into two intermediate frequencies (IFs),
of 50 MHz bandwidth each.
We used 3C~286 as the primary flux calibrator (assumed of constant flux density)
at all frequencies.
We observed SN~2001gd phase-referenced to J1310+32 (a nearby VLA phase-calibrator
source) at all bands, except at 8.4~GHz, for which we used our VLBI amplitude
calibrator, J1317+34, as the phase-reference source.
J1310+32 and J1317+34 served also as the phase-reference sources for the 
system calibration.
The flux densities reported here were obtained by combining the data from
both IFs at each frequency band.
As with our VLBI data, we used standard calibration and hybrid mapping
techniques to obtain a VLA image of SN~2001gd and its host galaxy, NGC~5033,
at the observing frequency of 8.4 GHz
(see Fig.~\ref{fig,bp103_xvla}).

%
\begin{figure}
\begin{center}
\mbox{\hspace{-0.3cm}
\epsfxsize=8.5cm
\epsfysize=8.5cm
\epsfbox{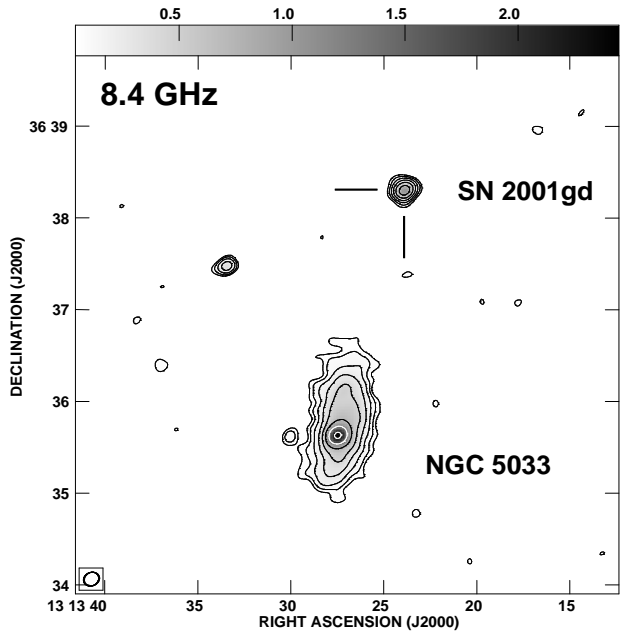}\hspace{0cm}}
\caption{
Hybrid image of the galaxy NGC\,5033 and its supernova SN\,2001gd
made at 8.4~GHz with the Very Large Array (VLA), from observations on
8 April 2003.
The contours are (3,$3\,\sqrt{3}$,9,...)$\times$ 16\,$\mu$Jy beam$^{-1}$,
the off-source rms flux density per beam.
The peak of brightness of the image corresponds to the nucleus of NGC\,5033
and is 2.44 mJy beam$^{-1}$.
The supernova is the bright point-like source northwards of the
nucleus of NGC\,5033.
The principal dimensions (full width at half maximum) of the restoring beam
are 10\farcs1 $\times$8\farcs7, with the beam's major axis
oriented along a position angle  of -61\degr (see inset in the lower left 
hand corner).
1\arcsec\ in the image corresponds to a linear size of 64~pc.
}
\label{fig,bp103_xvla}
\end{center}
\end{figure}

\subsection{Two-epoch XMM-Newton X-ray observations}
\label{subsec23}

NGC\,5033 was observed with the
\emph{XMM-Newton Observatory} on
2 July 2001      (PI: Turner,   Obs.\ ID: 0112551301) and on
18 December 2002 (PI: Quataert, Obs.\ ID: 0094360501).
Both observations used the EPIC/MOS1, EPIC/MOS2, and EPIC/pn CCD
cameras (\citealt{struder01}; \citealt{turner01});
the two EPIC/MOS cameras were operated in the Prime Full-Window Mode,
while the EPIC/pn camera was operated in the Extended Prime Full-Window
Mode.
The useful exposure times were $\sim$7.6 ks and $\sim$11.6 ks for the EPIC/MOS
cameras and $\sim$4.0 ks and $\sim$10.0 ks for the EPIC/pn camera for the 
observations ID 0112551301 and 0094360501, respectively.
The observation ID 0112551301 used the Thin Filter for the EPIC/MOS
observations; the Medium Filter was used for all the other observations.

We retrieved the \emph{XMM-Newton} pipeline products from the
\emph{XMM-Newton} Science Archive\footnote{
The \emph{XMM-Newton} Science Archive is supported by ESA and can be
accessed at http://xmm.vilspa.esa.es}.
Initial processing and analysis were performed using the
\emph{XMM-Newton}
Science Analysis Software (SAS ver.\ 6.0.0) and the calibration files
from
the Calibration Access Layer available on 13 April 2004.
Subsequent spectral analysis was carried out with XSPEC.

We examined the light curve of the EPIC/MOS and EPIC/pn in the energy
range above 10 keV to assess the effect of any high energy particle flux 
on the background level.
Although the light curve of the observation ID 0094360501 on 18 December 
2002 does not show any period of enhanced background, 
the observation ID 0112551301 on 2 July 2001 is severely
affected, with a background level 40 times higher than  
on the 18 December 2002 observation.
Finally, the data were filtered to remove poor event grades 
(most likely spurious X-ray events).

We extracted images from the EPIC/MOS and EPIC/pn observations in
different energy bands with pixel size of 1\arcsec.
The images extracted from the observation on 2 July 2001
do not show any X-ray source at the location of SN~2001gd. 
On the other hand, the image in the 0.25$-$2.5 keV band 
extracted from the observation on 18 December 2002 
(Fig.~\ref{fig,sn01gd_xmm}) does 
show a compact, soft X-ray source at the
location of SN~2001gd, with 19$\pm$5 counts in the 0.3--2.0 keV band.
In the energy band above 2 keV, there are only 2$\pm$2 counts, which
corresponds to a hardness ratio of 0.1$^{+0.2}_{-0.1}$ between the energy
band above 2 keV and the total \emph{XMM-Newton} energy band.
The facts that the pre-supernova observation yielded a non-detection and 
the post-supernova observation yielded a detection, coupled with
the spatial coincidence and soft X-ray spectrum imply that the 
X-ray source is SN~2001gd, and that the supernova explosion 
likely happened after 2 July 2001. 
%
\begin{figure}
\begin{center}
\mbox{\hspace{-0.3cm}
\epsfxsize=8.5cm
\epsfysize=8.5cm
\epsfbox{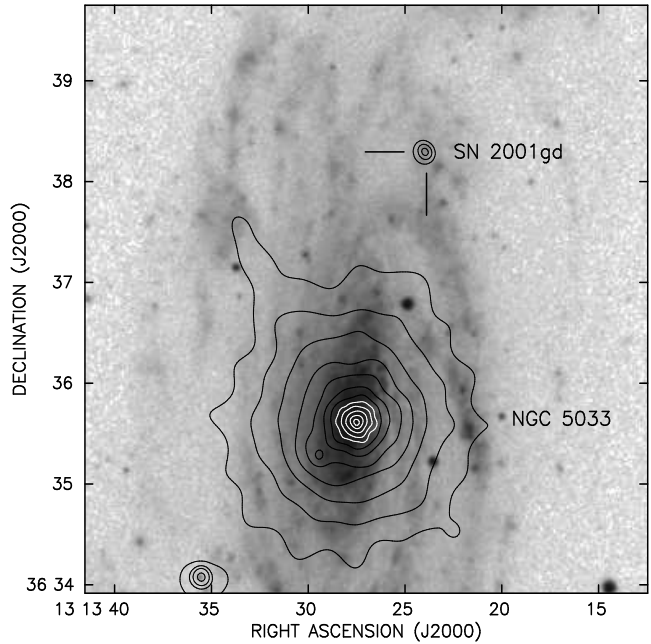}\hspace{0cm}}
\caption{
Contours of soft 0.25$-$2.5 keV X-rays from SN~2001gd
overlaid on a Second-Epoch Digitized Sky Survey 
(DSS-2) Blue plate of NGC~5033.
The X-ray image was extracted from the
18 December 2002 \emph{XMM-Newton} observation of NGC\,5033
(Obs.\ ID 0094360501).
The EPIC/MOS and EPIC/pn images have been combined to obtain a higher
S/N image.
The raw EPIC image has been further adaptively smoothed using Gaussian
profiles with FWHM ranging from 1\farcs5 to 9\arcsec.
The three contours in the image for SN~2001gd are drawn at 
(3, 8, and 13)$\times$0.15 cnts/arcsec$^2$, 
the image off-source rms.
}
\label{fig,sn01gd_xmm}
\end{center}
\end{figure}

%
\begin{figure}
\begin{center}
\vspace{-1.0cm}
\mbox{\hspace{-1.5cm}
\resizebox{11.0cm}{!}{\rotatebox{0}{\includegraphics{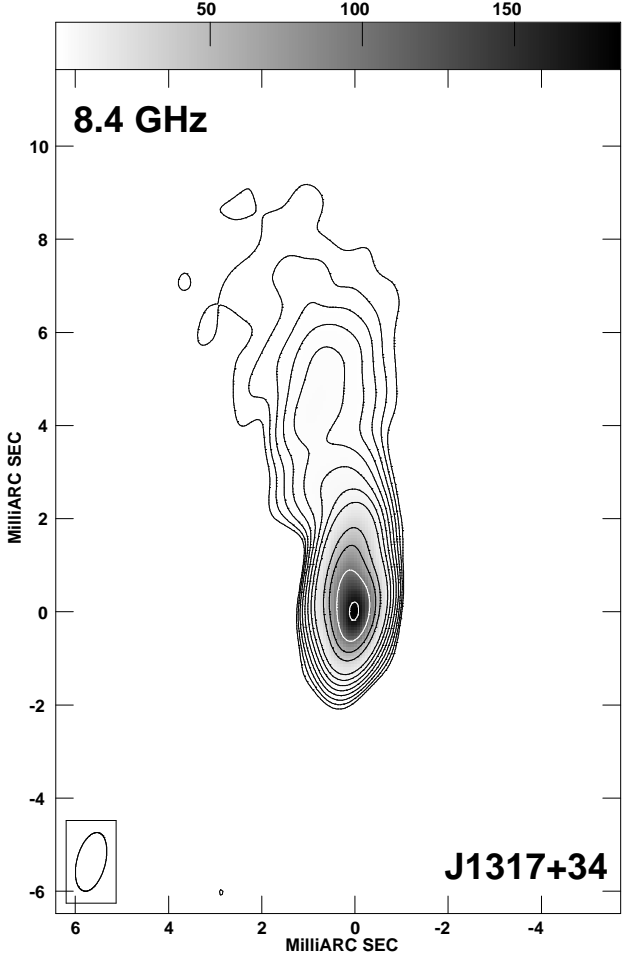}}}
}
\caption[]{
8.4~GHz Very-Long-Baseline Interferometry (VLBI) image of J1317+34
on 26 June 2002.
The contours are (3,$3\,\sqrt{3}$,9,...)$\times$80\,$\mu$Jy beam$^{-1}$,
the off-source rms flux density per beam.
The peak of brightness of the image is 184 mJy beam$^{-1}$ and
the restoring beam (bottom left in the image) is $1.29$ mas $\times 0.61$ mas
at a position angle of -14$\fdg$3.
The total flux density recovered in the image is $\sim0.33$\,Jy.
North is up and east is left.
One milliarcsecond in the image corresponds to a linear size of 8.5~pc.
}
\label{j1317_bp100}
\end{center}
\end{figure}

\section{VLBI images of SN~2001gd}
\label{sec,vlbi-images}

We used phase-reference techniques at both VLBI epochs to obtain images of
SN~2001gd. The use of phase-referencing allows detection and imaging of very
faint sources like supernovae, as it effectively increases
the coherence time from minutes to hours (e.g., \citealt{bc95}).
In our case, we made phase-reference observations of our target source,
SN\,2001gd, with respect to the relatively strong, nearby QSO~J1317+34,
which is $2\degs 22\amin$ distant.

We display in Fig.~\ref{j1317_bp100} our 8.4\,GHz VLBI image of J1317+34 on
26 June 2002.
The source shows a one-sided core-jet structure at milliarcsecond (mas) scales, 
with the jet extending northwards to $\sim9$\,mas.
As mentioned in Sect.~\ref{subsec,vlbi-obs}, we subtracted at each
epoch the phase contribution of 
J1317+34 (due to its extended structure) from the fringe phase,
and used the position of its phase-center (essentially coincident
with the peak of the brightness distribution), as our reference point for 
phase-reference imaging.

Figure~\ref{fig,sn01gd_vlbi} shows our VLBI images of
SN~2001gd, which showed no structure at both epochs. 
The coordinates of the supernova are offset from those provided
by the VLA \citep{stockdale02}, and used as {\it a priori} VLBI coordinates 
in the correlator, for the supernova position, 
by a mere $\sim$3 mas in right ascension 
and $\sim$1 mas in declination.
Since the standard error of the VLA position provided by \citet{stockdale02}
was 0.2 arcsec in each coordinate, the (small) differences between
the VLBI and VLA determined coordinates would be a highly improbable event.
It is more likely that the VLA uncertainties quoted by \citet{stockdale02} 
were far too conservative.
Indeed, a reanalysis of the 8.4 GHz VLA observations of \citet{stockdale02} 
results in 1~$\sigma$ uncertainties of 0.015--0.0020 arcsec (C. Stockdale, private communication) 
for the VLA position of SN~2001gd, about a factor of ten smaller than the 
uncertainties quoted in \citet{stockdale02}.

\begin{figure}
\begin{center}
\vspace{-0.25cm}
\mbox{\hspace{-0.3cm}
\epsfxsize=8.5cm
\epsfysize=8.5cm
\epsfbox{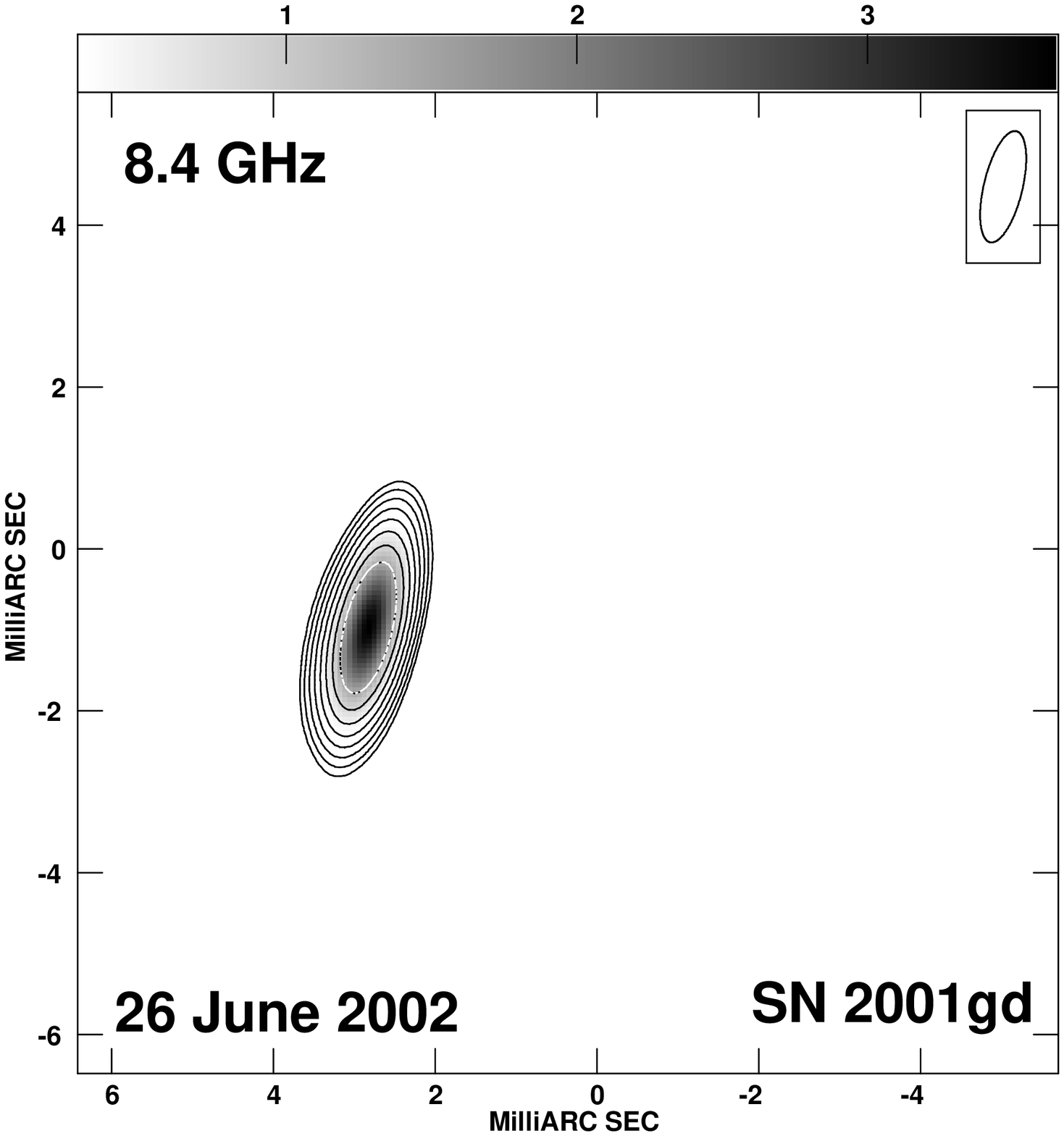}\hspace{0cm}
}
\mbox{\hspace{-0.3cm}
\epsfxsize=8.5cm
\epsfysize=8.5cm
\epsfbox{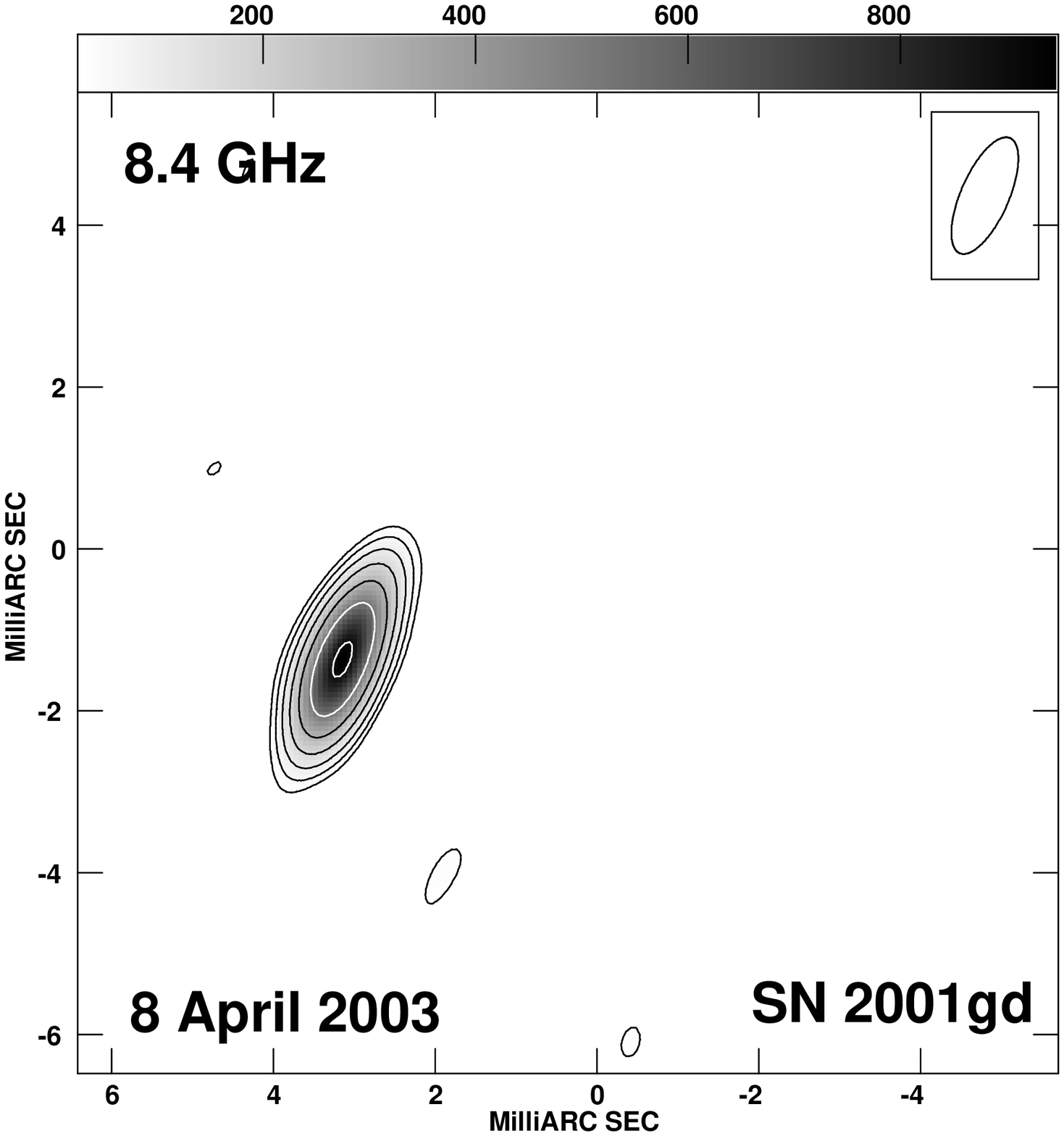}\hspace{0cm}
}
\caption{
{\em Top}: Very-Long-Baseline Interferometry (VLBI) image of SN\,2001gd
on 26 June 2002.
Contours are drawn at (3,$3\,\sqrt{3}$,9,...)$\times$10\,$\mu$Jy beam$^{-1}$,
the off-source rms.
The restoring beam (top right) is $1.41 \times 0.48$ mas$^2$
at a position angle of -13$\fdg$4. 
The peak of brightness is 3.64 mJy beam$^{-1}$, and the total  
flux density recovered for the supernova is 3.83\,mJy.
{\em Bottom}:
Very-Long-Baseline Interferometry (VLBI) image of SN\,2001gd on 8 April 2003.
Contours are drawn at (3,$3\,\sqrt{3}$,9,...)$\times$11\,$\mu$Jy beam$^{-1}$,
the off-source rms.
The restoring beam (top right) is $1.55 \times 0.59$ mas$^2$
at a position angle of -23$\fdg$7. 
The peak of brightness is 944\,$\mu$Jy beam$^{-1}$, and 
the flux density recovered for the supernova is 1.02\,mJy. 
In both panels, north is up and east is left; the origins are taken 
from the VLA observations. 
}
\label{fig,sn01gd_vlbi}
\end{center}
\end{figure}

\section{Results and Discussion}
\label{sec,discussion}

\subsection{Radio light curve and spectral behaviour}
\label{sec,light-curve}

Simultaneous multiwavelength radio observations of supernovae
are important to help characterize the relevant physical 
processes taking place. 
Since we were granted VLA observing time for 
our second VLBI epoch on 8 April 2003 
we were able to characterize the radio spectral energy distribution of SN~2001gd.
The flux density measurement error given for the VLA measurements in 
Table~\ref{tab,vla_log} represents one statistical standard deviation, 
and is a combination of the off-source rms in the image 
and a fractional error, $\epsilon$, included to account for the normal inaccuracy of 
VLA flux density calibration and possible deviations of the primary calibrator 
from an absolute flux density scale: 
the final errors, $\sigma_f$, as listed in Table~\ref{tab,vla_log}, 
are taken as  
$\sigma_f^2 = (\epsilon\,S_0)^2 + \sigma_0^2$
(see, e.g., \citealt{stockdale03}), 
where $S_0$ is the measured flux density, $\sigma_0$ is the off-source rms at a given
frequency, and $\epsilon$=0.10 at 1.4~GHz, 0.05 at 4.9~GHz, 
0.05 at 8.4~GHz, 0.075 at 15.0~GHz, and 0.10 at 22.5 and 43.3~GHz.
We fit our data using an implementation of the nonlinear weighted-least-squares 
Marquardt-Levenberg algorithm.
We plot in Fig.~\ref{fig,spectra} two model fits to the data in Table~\ref{tab,vla_log}.
The solid line corresponds to a fit to the data with a power-law spectrum
that is partially suppressed by free-free absorption from a homogeneous
screen of ionised gas 
[$S_\nu\,=\,S_1\,\nu^\alpha\,{\rm exp}(-\tau_{\rm ff, \nu})]$, where
$S_1$ is the flux density at 1\,GHz, in mJy; $\nu$ is the observing frequency, in 
GHz; and $\tau_{\rm ff, \nu}=A\,\nu^{-2.1}$ is the free-free optical depth
at the observing frequency $\nu$. 
The dashed-dotted line corresponds to a  pure synchrotron spectrum fit. 
For our first model, we obtain the following 
weighted-least-squares (``best-fit'') values 
for the parameters:
$S_1$=(9.1$\pm1.6)$~mJy, $\alpha$=-1.0$\pm0.1$, $A$=(1.0$\pm$0.3)~GHz$^{2.1}$. 
(The quoted uncertainties correspond to 1$\sigma$.)
The observed spectral index is typical of the optically thin phase of  
Type II supernovae.
The corresponding optical depth at 8.4\,GHz is 
$\tau_{\rm ff, 8.4}$=(1.1$\pm$0.3)\,$\times\,10^{-2}$. 
This estimate can be used to constrain the electron temperature, 
$T_e$=$10^5\,T_5$ K, 
and mass-loss rate parameter of the supernova progenitor, $\mdot/v_w$. 
(Each letter symbol with a subscript indicates the value of the corresponding 
quantity in units of 10 to the power given by the subscript; for example, $T_5$ above
denotes $T_e$ in units of 10$^5$ K.)
Indeed, for a steady presupernova mass loss rate, 
$\mdot$=10$^{-5}\,\mdot_{-5}\,\msunyr$, and wind speed, 
$v_w$=10\,$v_{10}\,\kms$, 
the free-free absorption optical depth 
of the unshocked ionised gas is (\citealt{lf88})

\begin{equation}
\tau_{\rm ff, \nu} \approx 
0.17\,
g_{\rm ff} \,\mu \, 
T_5^{-3/2}\,
\nu^{-2}\,
\mdot^2_{-5}\,v_{10}^{-2}\,
r_{16}^{-3}
\label{eq,tau}
\end{equation}

\noindent
where $\nu$ is the frequency in GHz, $r=10^{16}\,r_{16}$\,cm,
is the radius of the circumstellar shock, 
$g_{\rm ff}$ is the free-free Gaunt factor, and 
$\mu = [1+2\,n({\rm He})/n({\rm H})]/[1+4\,n({\rm He})/n({\rm H})]$
is the mean molecular weight for the presupernova wind. 
For solar abundances, $n({\rm He})/n({\rm H})$=0.1 and $\mu$=0.86, 
while for $n({\rm He})/n({\rm H})$=1, as may have been the case for
SN~1993J \citep{baron94}, $\mu$=0.6.  
Using $r_{16}=3.6$ (see Sect.~\ref{sec,expansion} below) and 
evaluating $g_{\rm ff}$ at 8.4\,GHz,  
we obtain 
$\tau_{\rm ff, \nu} \approx 2.1\EE{-4}\,(\mu/0.6)\,
T_5^{-3/2}\,
\mdot_{-5}^2\,
v_{10}^{-2}$. 
\citet{flc96} and \citet{fb98} showed 
that $T_5\,\approx\,$2 and $(\mdot_{-5}/v_{10})\approx\,$5 
were needed to explain 
the X-ray and radio emission of SN~1993J.
Using the above values of $T_5$ and $(\mdot_{-5}/v_{10})$  
for SN~2001gd on 8 April 2003, we 
obtain from Equation~(\ref{eq,tau})
$ \tau_{\rm ff, 8.4} \approx 2\EE{-3}$, 
which is a factor of three 
too low to fit the lower end of our
value for $\tau_{\rm ff, 8.4}$. 
We explored the space of possibilities within 2\,$\sigma$ of
our best-fit values for the parameters $S_1, \alpha$, and $A$ (see above). 
We found very good matches for $\tau_{\rm ff, 8.4}$ for the pairs of values 
($T_e$=$3\EE{4}$\,K; $\mdot$=$2.5\EE{-5}\,\msunyr$) and 
($T_e$=$2\EE{5}$\,K; $\mdot$=$1\EE{-4}\,\msunyr$).  
Unfortunately, the coupling of $T_e$ and $(\mdot/v_w)$ prevents us from favouring
one pair of values over any other solely based on the radio observations, 
so we limit ourselves to pointing out
that the values of $T_e$ and $\mdot$ for SN~2001gd
are likely to be in the range $T_e$=$(3-20)\EE{4}$\,K, 
$\mdot$=$(2-10)\EE{-5}\,\msunyr$.
As we will see in Sect.~\ref{sec,xrays}, the available X-ray data 
favour models with $\mdot\,\lsim\,5\EE{-5}\,\msunyr$, which would imply electron
temperatures of $T_e\,\lsim\,7\EE{4}$\,K to satisfy our radio observations.

\begin{table*}
 \caption[]{Results from VLA observations on 8 April 2003}
  \label{tab,vla_log}
$$
 \begin{array}{lrrrrrr}
   \hline\noalign{\smallskip}
  \multicolumn{1}{l}{\rm Source\, Name} & 
  \multicolumn{6}{c}{\rm Flux\, density\,(mJy) } \\ 
        \cline{2-7}\noalign{\smallskip} 
&
\multicolumn{1}{c}{43.3\, \rm GHz} &
\multicolumn{1}{c}{22.5\, \rm GHz} &
\multicolumn{1}{c}{15.0\, \rm GHz} &
\multicolumn{1}{c}{8.4\, \rm GHz} &
\multicolumn{1}{c}{4.9\, \rm GHz} &
\multicolumn{1}{c}{1.4\, \rm GHz} \\ 
   \hline\noalign{\smallskip}
\multicolumn{1}{l}{\rm SN~2001gd} &
\multicolumn{1}{c}{\lsim 0.30^\dagger } &
\multicolumn{1}{c}{0.39 \pm 0.12} &
\multicolumn{1}{c}{0.63 \pm 0.13 } &
\multicolumn{1}{c}{1.02 \pm 0.05} &
\multicolumn{1}{c}{1.85 \pm 0.24} &
\multicolumn{1}{c}{3.89 \pm 0.40} \\ 
\multicolumn{1}{l}{\rm 1310+323^{(1)}} &
\multicolumn{1}{c}{2584 \pm 258} &
\multicolumn{1}{c}{2911 \pm 291}  & 
\multicolumn{1}{c}{2397 \pm 178 }  &
\multicolumn{1}{c}{ - }  &
\multicolumn{1}{c}{1656 \pm 83}  &
\multicolumn{1}{c}{1535 \pm 154 } \\ 
\multicolumn{1}{l}{\rm J1317+34^{(2)}} &
\multicolumn{1}{c}{ - } &
\multicolumn{1}{c}{ - } &
\multicolumn{1}{c}{ - } &
\multicolumn{1}{c}{ 384 \pm 19 } &
\multicolumn{1}{c}{ - } &
\multicolumn{1}{c}{ - } \\ 
\hline\noalign{\smallskip}
 \end{array}
$$
\begin{list}{}{}
\item[] {\rm $^{(1)}$ Phase and secondary flux density calibrator at all 
frequencies, except at 8.4~GHz.
$^{(2)}$ Phase and secondary flux density calibrator at 8.4~GHz.
$^\dagger$ The quoted flux density at 43.3 GHz is an upper limit, corresponding to 
three times the off-source rms noise in the image.
The quoted uncertainty for each flux density value corresponds to 
1$\sigma$ (see Sect.~\ref{sec,light-curve} for details).
 }
\end{list}
\end{table*}

%
\begin{figure}
\begin{center}
\resizebox{8.6 cm}{!}{\rotatebox{-90}{\includegraphics{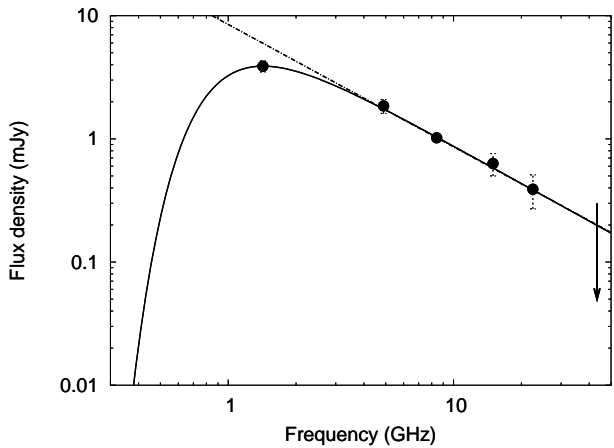}}}
\caption[]{
Model fits to the radio spectrum of SN~2001gd on 8 April 2003.
The solid line corresponds to a synchrotron spectrum fit, partially suppressed by
free-free absorption, whilst the dashed-dotted line corresponds to a pure 
synchrotron
spectrum fit for all frequencies, except 1.4 GHz.
The error bars denote $\pm 1\sigma$ values.
}
\label{fig,spectra}
\end{center}
\end{figure}

%
\begin{figure}
\begin{center}
\vspace{-1.0cm}
\mbox{\hspace{-2.4cm} 
\resizebox{15.5cm}{!}{\rotatebox{0}{\includegraphics{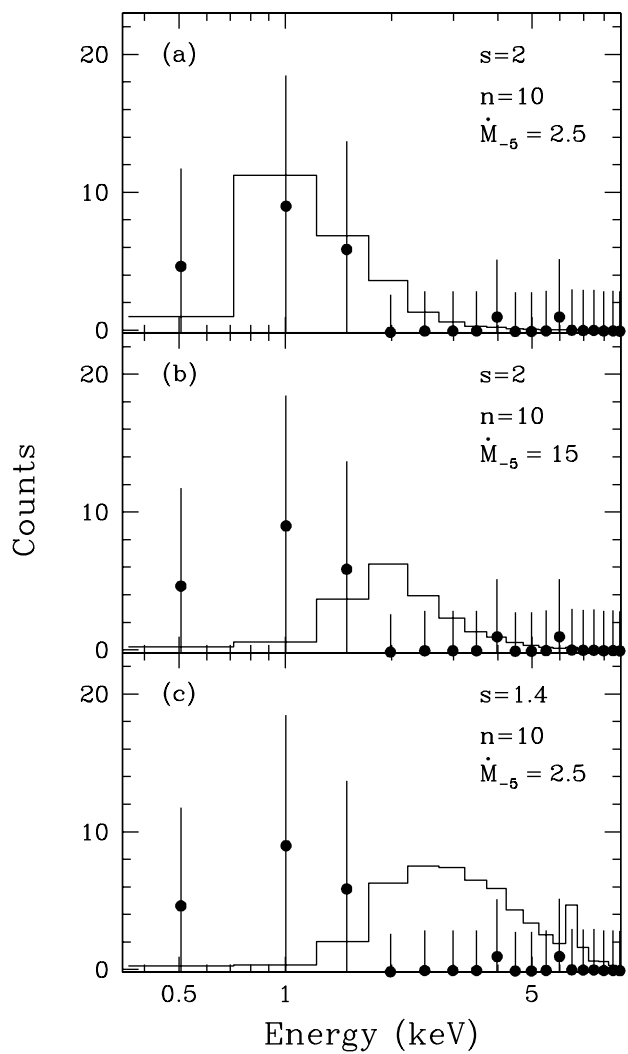}}}
}
\caption{
{\it XMM-Newton} EPIC/pn spectrum of SN~2001gd on 18 December 2002, 
overplotted by thin plasma X-ray emission
models for a supernova radiative reverse shock \citep{flc96}. 
The different sets of parameters are displayed in the corresponding panel.
The uncertainties shown are one statistical standard deviation.
The energy bin-width is 0.5 keV; therefore, 1 cnt in the spectrum corresponds to 
$\sim 2.0\EE{-4}$ cnt s$^{-1}$ keV$^{-1}$.
}
\label{fig,xspec}
\end{center}
\end{figure}

\subsection{X-ray emission from SN~2001gd}
\label{sec,xrays}

Models for the circumstellar interaction 
around supernovae (e.g., \citealt{flc96}) predict the existence of two 
components in X-rays: a soft component associated with the reverse shock
of the supernova, $T_{\rm rs}$, 
and a hard component associated with the circumstellar shock, $T_{\rm cs}$, 
e.g., SN~1995N \citep{mucciarelli04}.
The 18 December 2002 ($t-t_0$=472~d) \emph{XMM-Newton} observations of NGC~5033
can be used to investigate the circumstellar interaction around SN~2001gd and 
to estimate its X-ray luminosity. 
However, the spectrum has too small a number of counts to allow spectral fitting. 
Instead, we compared the count rate and hardness ratio of the 
observed \emph{XMM-Newton} EPIC/pn spectrum with these implied from 
the analytical models for X-ray supernovae by \citet{flc96}.
Figure~\ref{fig,xspec} shows the {\it XMM-Newton} spectrum of SN~2001gd on 18 
December 2002, overplotted by several optically thin plasma X-ray emission models.
According to these models, both column density absorption and electron temperature
are sensitive to the circumstellar density
profile ($\rho_{\rm csm} \sim r^{-s}$), 
ejecta density profile ($\rho_{\rm ej} \sim r^{-n}$), 
shock speed ($v_{\rm sh}=10^4\,v_4\,\kms$), and
mass-loss rate ($\mdot$=10$^{-5}\,\mdot_{-5}\,\msunyr$). 
As for the expansion velocity, 
H$_\alpha$ spectra taken on December 2001 and January 2002, 
indicate $v_{\rm HWHM} \approx 7800 \pm 1000 \kms$ (T. Matheson, private communication).  
Considering a likely supernova deceleration from this date to the time when the 
X-ray observations were obtained, 
we adopted a supernova shock speed of $v_4$=0.6 on 18 December 2002. 
We also kept fixed the presupernova wind speed at $v_{10}$=1, 
while we varied $s$, $n$, and $\mdot$ to investigate their effects on the X-ray emission.
We note that the analytical models of \citet{flc96} do not predict the exact amount 
of X-ray luminosity, but the reverse shock total luminosity, $L_{rs}$.  
A close inspection of their results indicates that the X-ray luminosity is
$\sim$20\% of $L_{rs}$, and thus   
we adopted an X-ray luminosity of this fraction of $L_{rs}$
to estimate the \emph{XMM-Newton} EPIC/pn count rate.

The supernova model that shows hardness ratio and X-ray luminosity
most consistent with the \emph{XMM-Newton} spectrum of SN~2001gd 
on 18 December 2002 has $k\,T_{\rm rs}\approx\,$1.1 keV, 
$s$=2, $n$=10, and $\mdot_{-5}$=2.5. 
For illustration purposes, we compare in 
Fig.~\ref{fig,xspec} the spectral shape of this model (top panel) with the 
observed X-ray spectrum.
The above parameters agree well with what is expected for a supernova radiative reverse 
shock \citep{flc96}. 
Our value of $\mdot_{-5}$=2.5 for the mass-loss rate of SN~2001gd 
from the X-ray observations may be compared with the value of 
$\mdot_{-5}$=3 obtained by \citet{stockdale03} from their radio 
light curve modelling, and with our own estimates in 
Sect.~\ref{sec,light-curve}. 
The corresponding column density of the cooling shell between the reverse
shock and the contact surface is $N_{\rm cool}=5.5\EE{21}$\,cm$^{-2}$, and
that of the circumstellar shock is $N_{\rm cs}=1.6\EE{21}$\,cm$^{-2}$. 
We note that the supernova shock speed, $v_{\rm sh}$, 
and the ejecta density profile, $n$, are coupled, so that 
different combinations within a certain range of these parameters 
reproduce the observed hardness ratio and count rate.
Models with supernova shock speeds as high as
$v_{\rm 4} \approx 1.2$, would require $n \approx 14$ to reproduce the observed
hardness ratio, but result in unreasonably high \emph{XMM-Newton} EPIC/pn count rates. 
On the other hand, 
models with supernova shock speeds as low as 
$v_{\rm 4} \approx 0.2$, would require $n \gsim 5$ to reproduce the observed
hardness ratio, but result in too low \emph{XMM-Newton} EPIC/pn count rates. 
We also explored models with mass loss rates and circumstellar density profiles 
in the ranges 
$0.1\le \mdot_{-5} \le 30$ and $1.1 \le s \le 2.9$, respectively. 
Models with $\mdot_{-5} \lsim 0.5$, while predicting consistent hardness ratios, 
are unreasonable as their cooling times are much greater than the age of SN~2001gd.
Models with $s \gsim 2.4$ predict significantly lower \emph{XMM-Newton} 
EPIC/pn count rates than observed. 
Models with $\mdot_{-5} \gsim 15$ (middle panel in Fig.~\ref{fig,xspec})
or  $s \lsim 1.4$ (bottom panel in Fig.~\ref{fig,xspec})
have hardness ratios between the energy band 
above 2 keV and the total \emph{XMM-Newton} energy band that are 
$\gsim 0.6$, well above the observed hardness ratio of $0.1^{+0.2}_{-0.1}$ (see Sect.~\ref{subsec23}). 
Therefore, models with $\mdot_{-5} \lsim 0.5$ or  $\mdot_{-5} \gsim 15$,  
and models with $s \lsim 1.4$ or $s \gsim 2.4$ seem less likely to describe the X-ray emission from SN~2001gd 
than the model shown in the top panel of Fig.~\ref{fig,xspec}.

At the adopted distance of 13.1~Mpc, the unabsorbed flux from SN~2001gd 
is $F_{\rm X} = (7\pm 2)\EE{-14}$\,ergs~cm$^{-2}$\,s$^{-1}$ in the 0.3--5 keV band,  
which corresponds to a luminosity of 
$L_{\rm X}=4\,\pi D^2 F_{\rm X}$= (1.4$\pm$0.4)$\times$10$^{39} D^2_{13}$ 
ergs~s$^{-1}$,  
where $D$=$13.1\,D_{13}\,{\rm Mpc}$ is the distance to the supernova. 
The above results are in agreement with expectations from X-ray emission from 
Type II supernovae (e.g., \citealt{flc96}), and seem to favour a 
relatively shallow ejecta density profile and a mass-loss rate about
a factor of two lower than for SN~1993J.

The 2 July 2001 \emph{XMM-Newton} observations of SN~2001gd did not 
yield detectable X-ray emission, 
but can be used to derive an upper limit of the isotropic X-ray
luminosity of SN~2001gd at that time.
The 3$\sigma$ upper limit on this luminosity  is
$ L_{\rm X}\,=\,7.4 \times 10^{38}$ erg~s$^{-1}$.
Note, however, that this upper limit to the X-ray luminosity
was obtained with a significantly shorter exposure time, and   
with a much higher background X-ray emission level, compared  to 
that for the observations on 18 December 2002.

Finally, the expected monochromatic flux density at 4.9\,GHz 
on 18 December 2002 ($t-t_0$=472~d)
was $\approx$\,(2.5$\pm$0.3)\,mJy. 
Using $\alpha=-1.0$ from our best-fit VLA spectrum on 8 April 2003, 
we obtain an isotropic radio luminosity of 
$L_{\rm R}\,=\,(8\,\pm\,2)\EE{36}\,D_{13}^2$\ergs between 1.4 and 43~GHz. 
The values of $L_{\rm X}$ and $L_{\rm R}$ put SN~2001gd close to SN~1993J
\citep{chevalier03}
regarding radio and X-ray luminosities, in agreement
with the fact that SN~2001gd displayed a very similar behaviour 
to SN~1993J in the optical band.

\subsection{Expansion of SN~2001gd}
\label{sec,expansion}

The VLBI images of SN~2001gd at 8.4~GHz (Fig.~\ref{fig,sn01gd_vlbi}) 
show a compact, unresolved radio emitting structure.
Therefore, we resorted to model fitting
to estimate the supernova angular diameter, $\theta$, at each epoch.
For this purpose, we used the model fitting procedure 
included in the Caltech DIFMAP package, which fits a given model 
directly to the real and imaginary parts of the observed interferometric visibilities, 
using the Levenberg-Marquardt non-linear least squares minimization technique.
Inspired by the findings for SN~1993J (e.g., \citealt{marcaide95a, marcaide95b, 
marcaide97, bartel00}), we 
assumed spherical symmetry in our model fits, and 
considered the following models:
(i) an optically thin sphere;
(ii) a uniformly bright, circular disk; and
(iii) an optically thin shell of width 25~\% its outer shell radius (e.g. \citealt{marscher85}).
Table 2 summarizes our results, where the uncertainties (1$\sigma$)
represent one statistical standard deviation plus an estimate of modelling errors. 
All three models give similarly good fits to the ($u,v$) data 
($\chi^2_{\rm red} = 0.8-0.9$), 
so in principle we cannot rule out any of them.
Our angular diameter estimates nominally suggest a relatively strong deceleration
in the expansion of SN~2001gd,  
but values of the deceleration parameter $m$ ($\theta \propto t^m$) between 0 and 1 are all 
within the uncertainties, and therefore we cannot draw any conclusion from our data.  

\begin{table}
\label{tab,sizes}
\centering
\begin{minipage}{84mm}
\caption{\protect{Angular diameter estimates of SN~2001gd}}
\begin{tabular} {lll} \hline
\multicolumn{1}{c}{Model} &
\multicolumn{2}{c}{Angular diameter ($\mu$\,as)} \\
	\cline{2-3}\noalign{\smallskip}
&
\multicolumn{1}{c}{2002.48}  & 
\multicolumn{1}{c}{2003.27}  \\  \hline 
{\sc optically thin sphere} & $390\pm80$ & $440\pm100$  \\ 
{\sc uniformly bright disk}   & $350\pm60$ & $410\pm80$  \\ 
{\sc optically thin shell$^\dagger$}  & $330\pm60$ & $370\pm80$  \\ 
\hline \\
\end{tabular}
\begin{list}{}{}
\item[] {
\rm $^\dagger$ The shell width is 25\% the outer radius.}
\end{list}
\end{minipage}
\end{table}

\subsection{Energy budget and magnetic field in SN~2001gd}
\label{emin-bmin}

The high brightness temperatures inferred from our VLBI observations for SN~2001gd
[(5.3$\pm$1.2)$\times\,10^8$\,K on 26 June 2002,
and (1.1$\pm$0.3)$\times\,10^8$\,K
on 8 April 2003], indicate a non-thermal, 
synchrotron origin for the radio emission from SN~2001gd. 
We can estimate a minimum total energy in relativistic particles and fields, 
and an equipartition magnetic field for SN~2001gd, 
by assuming equipartition between fields and particles.
The minimum total energy and its equipartition magnetic field are
\citep{pachol70}:

\begin{equation}
E_{\rm min} = c_{13}\, (1 + \psi)^{4/7}\, \phi^{3/7}\, R^{9/7}\, L_{\rm R}^{4/7}
\label{eq,emin}
\end{equation}
\begin{equation}
B_{\rm min} = (4.5\,c_{12}/\phi)^{2/7}\,(1 + \psi)^{2/7}\,  R^{-6/7}\, L_{\rm R}^{2/7}
\label{eq,bmin}
\end{equation}

\noindent
where $L_{\rm R}$ is the radio luminosity of the source;
$R$, its linear radius;
$c_{12}$ and $c_{13}$, slowly-varying functions of the spectral index, 
$\alpha$ \citep{pachol70};
$\phi$, the filling factor of fields and particles; and
$\psi$ the ratio of the (total) heavy particle energy to the electron energy.
This ratio depends on the mechanism that generates the relativistic electrons 
and ranges from $\psi \approx 1$ to $\psi = m_p/m_ e \approx 2000$, where $m_p$ and $m_e$
are the proton and electron mass, respectively.

From our VLA observations on 8 April 2003, 
we determined a spectral index $\alpha =-1.0 \pm 0.1$ 
($S_\nu \propto \nu^{\alpha}$; see Sect.~\ref{sec,light-curve}) 
between 1.4 and 43\,GHz, and 
$S_{8.4 \rm GHz} = 1.02\pm0.05$~mJy. With these values, we obtain 
$L_{\rm R}\,=\,(6.0\pm0.3) \times 10^{36}\,D_{13}^2$\,\ergs\ 
between 1.4 and 43\,GHz. 
We used $\phi$=0.6\,$\phi_{0.6}$ in our calculations, 
which corresponds approximately to the filling factor for the shell model in Table 2.
In any case, the estimates of $E_{\rm min}$ and $B_{\rm min}$ are relatively insensitive
to the precise value of $\phi$. 
The radius of SN~2001gd on 8 April 2003 is, for the shell model,
$R = 3.6\EE{16}\,D_{13}\,\theta_{185}$\,cm, where $\theta_{185}$ is the angular radius
of the supernova in units of 185\,mas.  
With these values, we get from Eq.~(\ref{eq,emin}) and (\ref{eq,bmin})

\begin{displaymath}
 E_{\rm min}\,\approx\,1.8\EE{46} \,(1 + \psi)^{4/7}\, 
   \phi_{0.6}^{3/7}\,
   \theta_{185}^{9/7}\,   
   D_{13}^{17/7}\,  {\rm ergs}
\end{displaymath}

\begin{displaymath}
 B_{\rm min}\,\approx\, 40\,(1 + \psi)^{2/7}\, 
   \phi_{0.6}^{-2/7}\, 
   \theta_{185}^{-6/7}\,   
   D_{13}^{-2/7}  {\rm mG}
\end{displaymath}

\noindent
Since $1 \lsim \psi \lsim 2000$,
$E_{\rm min}$ is in the range  $(2.6\EE{46}-1.4\EE{48})$\,erg,
and the equipartition magnetic field in the range (50--350)\,mG.
Thus, for an energy of 10$^{51}$\erg\ for the explosion of SN~2001gd, the fraction
of energy necessary to power the radio emission of SN~2001gd is quite modest.
The upper range of the equipartition magnetic field is lower than, 
but not far from, the lowest part of 
the range of values obtained for SN~1993J at similar radii
($B\,\approx$\,1.7~G, \citealt{fb98}; $B\,\approx$\,0.5~G, \citealt{mapt01}).

Since the kinetic energy density
in the wind is likely larger than the circumstellar magnetic field energy density
(e.g., \citealt{fb98}),
we have
$\rho_{\rm csm}\,v^2_{\rm w}/2 \gsim B_{\rm csm}^2/8\pi$, where
$\rho_{\rm csm}$ and $B_{\rm csm}$ are typical of the density and magnetic field in the
circumstellar medium, respectively.
By assuming a standard wind density profile ($\rho \propto r^{-2}$), we can write
the above expression as 

\begin{equation}
 B_{\rm csm} \lsim (\mdot\,v_w)^{1/2}\,r^{-1}\,\approx\, 2.5\,(\mdot_{-5}\,v_{10})^{1/2}\, r_{16}^{-1}\,{\rm mG} 
\label{eq,bfield}
\end{equation}

\noindent
where $r_{\rm 16} = r/10^{16}$\,cm. 
For $r_{\rm 16}=3.6$ and $\mdot_{-5} = 2.5$,
we obtain $B_{\rm csm} \lsim 1$\,mG, which is a factor about 50 to 350 times smaller 
than the equipartition field, 
and shows that the magnetic field inferred
for SN~2001gd cannot originate solely by compression 
of the existing circumstellar magnetic field, which would increase the
field only by a factor of four (e.g., \citealt{dyson})
Large amplification factors of the magnetic field have also been found 
for other radio SNe, e.g., SN1993J 
(amplification factors of a few hundred; \citealt{fb98}, \citealt{mapt01}), or SN~1986J 
(amplification factors in the range 40$-$300; \citealt{mapt02}). 
Thus, if equipartition between fields and particles exists, 
amplification mechanisms other than compression of the circumstellar magnetic field need to
be invoked, e.g., turbulent amplification \citep{chevalier82,cb95}, 
to explain the radio emission from supernovae.

\section{Summary}
\label{summary}

We have presented the results of the first two 8.4\,GHz very-long-baseline
interferometry observations of SN~2001gd in NGC~5033,
complemented with VLA and {\it XMM-Newton} observations.
We summarize our main results as follows:

\begin{itemize}
\item 
The radio structure of SN\,2001gd is not resolved, either on 2002.48
($t-t_0=295$\,days; $t_0$=3 September 2001) or on 2003.27
($t-t_0=582$\,days).  We used the interferometric visibility data to
estimate the angular sizes for the supernova, as well as constraints
on its expansion speed from optical measurements.  While our data
nominally suggest the possibility of a relatively strong deceleration
in the expansion of SN~2001gd ($m \ll 1$, $\theta \propto t^m$),
solutions with $m=1$ are also permitted, and therefore we cannot draw
any conclusion from these data.

\item
Our VLA data on 8 April 2003 can be well fit by an optically thin, synchrotron
spectrum ($\alpha = -1.0 \pm 0.1; S_\nu\, \propto\, \nu^{\alpha})$,
which is partially absorbed by thermal electrons.
The radio spectrum and the light curve at 1.4 GHz
indicate that the synchrotron turnover frequency is near 1\,GHz.
We obtain a supernova flux density of (1.02$\pm$0.05)\,mJy at the
observing frequency of 8.4 GHz.
At an adopted distance of 13.1~Mpc to NGC~5033, this flux density implies 
an isotropic radio luminosity of (6.0$\pm$0.3)$\times\,10^{36}$\ergs\ between
1.4 and 43\,GHz.
We also used our best-fit to the VLA radio spectrum to infer 
the most likely ranges for the electron temperature in the wind, $T_e$, 
and the mass-loss rate, $\mdot$. 
From our VLA observations on 8 April 2003, we 
find $T_e$=$(3-20)\EE{4}$\,K, and $\mdot$=$(2-10)\EE{-5}\,\msunyr$.
(Note the constraint on the combination of these values; 
see Sect. \ref{sec,light-curve} and \ref{sec,xrays}.)  

\item
By assuming equipartition between fields and particles,
we estimate a minimum total energy in relativistic particles and magnetic fields 
in the supernova shell of $E_{\rm min} \approx$(0.3--14)$\times\, 10^{47}\,$ erg,
which corresponds to an equipartion average magnetic field there
of $B_{\rm min} \approx$\,(50--350)\,mG.
We find that the average magnetic field in the circumstellar wind of SN\,2001gd
is $B_{\rm csm}\,\lsim 1$\,mG at a radius from the center of the supernova explosion 
of $\approx\,3.6\EE{16}$\,cm.
Since compression of this existing magnetic field by the supernova shock front 
could enhance it by only a factor of four, 
powerful amplification mechanisms must then be acting in SN\,2001gd,
to account for the magnetic fields responsible for the synchrotron radio emission.

\item We used {\it XMM-Newton} archival data to estimate the X-ray flux
of SN~2001gd on 2 July 2001 and 18 December 2002.
The data from the X-ray observations on 2 July 2001 are below 
the noise level, while
we detected SN~2001gd on 18 December 2002, with a
flux of $(7\pm2)\EE{-14}$\,erg~cm$^{-2}$\,s$^{-1}$ in 
the 0.3--5 keV band. 
The emission is chiefly from a soft component ($<$2 keV), 
associated with the reverse shock of the supernova. 
The corresponding isotropic X-ray luminosity is 
$L_{\rm X}$=(1.4$\pm$0.4)$\times$10$^{39}$ 
erg~s$^{-1}$ in the 0.3--5 keV band. 
The X-ray spectrum is consistent with expectations from X-ray emission from 
Type II supernovae (e.g., \citealt{flc96}), and suggests the 
interaction of a relatively shallow supernova ejecta density profile 
($\rho_{\rm ej} \sim r^{-n}; n \approx 10$),  with a standard circumstellar
wind density profile 
($\rho_{\rm csm} \sim r^{-s}; s\approx2.0$),  characterized by 
a presupernova mass-loss rate of $\mdot\,\approx \,2.5\EE{-5}\,\msunyr$ for SN~2001gd. 

\end{itemize}

SN~2001gd resembles
SN~1993J in its radio and X-ray emission, indicating
that a similar circumstellar interaction is taking place.
Since available radio supernovae for VLBI studies are so rare, 
it is important that the few of them that allow such studies be monitored if we 
are to better understand the radio supernova phenomenon.
Further VLBI observing epochs of SN~2001gd 
are necessary to trace the expansion history of the supernova, 
and to determine whether SN~2001gd is decelerating. 
However, the task is most challenging, as SN~2001gd appears to
fade very quickly in radio.  
We estimate that SN~2001gd will have a flux density 
at 8.4 GHz
of about 0.3--0.4\,mJy around March-April 2005. 
Therefore, the chances of a successful radio imaging
of SN~2001gd are small, and will necessarily require the use of an 
array including the world's most sensitive antennas.
 
\section*{Acknowledgments}
This research was partially funded by the grants AYA2001-2147-C02-01 
and AYA2002-00897 of the Spanish Ministerio de Ciencia y Tecnolog\'{\i}a.
MAPT and MAG are supported by the Spanish National programme Ram\'on y Cajal. 
KWW wishes to thank the Office of Naval Research for the 6.1 funding supporting 
this research.
We thank NRAO, the Max-Planck-Institut f\"ur Radioastronomie, and 
the Istituto di Radioastronomia for supporting our observing campaigns.
We thank an anonymous referee for useful comments and suggestions, which
significantly improved our manuscript.
NRAO is a facility of the USA National Science Foundation  
operated under cooperative agreement by Associated Universities, Inc.
We made use of the NASA Astrophysics Data System Abstract Service.
The Digitized Sky Surveys were produced at the Space Telescope 
Science Institute under U.S.\ Government grant NAG W-2166.  
The images of these surveys are based on photographic data 
obtained using the Oschin Schmidt Telescope on Palomar Mountain 
and the UK Schmidt Telescope.  
The plates were processed into the present compressed digital 
form with the permission of these institutions.

\label{lastpage}
\end{document}